\documentclass[10pt,a4paper]{article}
\usepackage[latin1]{inputenc}
\usepackage[T1]{fontenc}
\usepackage{amsmath}
\usepackage{amsfonts}
\usepackage{amssymb}
\usepackage{graphicx}

\begin{document}
\title{Light-Light scattering in Very Special Relativity Quantum Electrodynamics and Cosmic Anisotropies}

\author{J. Alfaro \\
	Facultad de F\'\i sica, Pontificia Universidad Cat\'olica de Chile,\\
	Casilla 306, Santiago 22, Chile.\\
	jalfaro@uc.cl}

\maketitle

\begin{abstract}
We consider scattering of  light by light  in Very Special Relativity (VSR) Quantum Electrodynamics(QED) with a non-zero photon mass. In order to preserve gauge invariance and Sim(2) symmetry we made use of a recently introduced infrared regularization of VSR theories. Additionally we show that all one loop diagrams with any number of photon external legs and zero fermion legs reduce to the standard QED result with the effective electron mass, as required by unitarity. It follows that the Euler-Heisenberg Lagrangian is  the same as in QED.
The total cross section of scattering of light by light exhibits tiny anisotropies that could be detected at cosmological scales. In particular they should be looked at the Cosmic Microwave Background(CMB) Radiation for low photon frequencies.
		
\end{abstract}
\section{Introduction}
The $SU (2)_L \times U (1)_R$ model of weak and electromagnetic
interactions with spontaneously symmetry breaking of the vacuum symmetries,
the Standard Model(SM), is being verified with a high precision at the LHC.
The discovery of the Higgs particle at the LHC in 2013 completed the particle
structure of the SM\cite{w2}.

But the SM cannot be the whole story. We know that the neutrino has a mass,
due to the discovery of neutrino oscillations\cite{Langacker}, whereas in the SM
neutrinos are massless.

Preservation of Lorentz symmetry forces the introduction of new particles and
interactions to take into account the neutrino mass as in the seesaw
mechanism\cite{mohapatra}.

Remarkably neither new interactions nor new particles beyond the SM have been
unveiled at the LHC.

Special relativity (SR) is valid at the largest energies available today \cite{PA}. However, its violation has been
contemplated as a possible evidence for Planck scale Physics, as some theories of
quantum gravity predicts it\cite{AMU}. For a review of this line of research, please see \cite{AC,JLM}.
Experiments and astrophysical observations are
used to put bounds upon the parameters describing these violations.
Basically, three scenarios have been studied: (1) Non-dynamical
tensor fields are introduced to determine preferred directions that break the
Lorentz symmetry. Some instances of this are the Myers-Pospelov model \cite{MP}
together with QED in a constant axial vector background \cite{AA}. (2) A second scenario is to assume spontaneous symmetry 
breaking (SSB) of the Lorentz symmetry as in the standard model extension of \cite{CK}, where such non-dynamical
tensor fields are assumed to arise from vacuum expectation values of some basic
fields belonging to a more fundamental theory. (3) Very Special Relativity(VSR).

It is possible to introduce a neutrino mass in the context of
VSR\cite{CG1}. The six parameter Lorentz group is reduced to a 3 parameters
subgroup ($Hom(2)$) or a 4 parameter subgroup ($Sim(2)$). Most
interesting for us is the $Sim(2)$ subgroup as a symmetry of Nature. This subgroup does not have invariant tensor fields besides the ones
that are invariant under the whole Lorentz group, implying that the dispersion
relations, time delay and all classical tests of SR are valid for this subgroup
also.It is defined by transforming a null vector $n_{\mu}$ at most by a multiplicative factor,
$n_{\mu} \rightarrow e^{\phi} n_{\mu}$, so ratios of scalar quantities containing
the same number of $n_{\mu}$ in the numerator as in the denominator are
$Sim(2)$ invariant, although  they are not Lorentz invariant. In
this context a new term in Dirac equation is allowed providing a VSR mass term for left handed
neutrinos{\cite{CG2}}.

VSR has been generalized to include supersymmetry
\cite{CZ}, curved spaces \cite{GG}, noncommutativity \cite{ST}, cosmological constant \cite{AV}, dark
matter \cite{AH}, cosmology \cite{CL}, Abelian gauge fields \cite{Cheon} and non-Abelian gauge fields \cite{AR}.

Some time ago, we built the Very Special Relativity Standard Model (VSRSM)\cite{ja1} .
It has the same symmetries and particles as in the SM. But we can describe
neutrino masses and neutrino oscillations using VSR. VSRSM is renormalizable
and unitarity is preserved. New processes such $\mu \rightarrow e + \gamma$
are predicted.

When computing radiative corrections in VSR models, a new problem appears.
There are infrared divergences due to non local terms of the form
$\frac{1}{n.p}$, where $p_{\mu}$ is a loop momentum. Earlier attempts to deal
with infrared divergences in VSR have been considered in \cite{plb,AUniverse,AS1,AS2}. To implement this,
Feynman rules have to be modified\cite{AUniverse}.

Recently \cite{jaren}, we have introduced a new infrared regularization that preserves
gauge invariance and $Sim(2)$ symmetry, using the standard Feynman
rules. It is based in Mandelstam-Leibbrandt (ML)\cite{Mandelstam,Leibbrandt} regularization  implemented
in \cite{AML}. ML introduces an additional null vector $\bar{n}_{\mu}$, $n.
\bar{n} = 1$. ML preserves gauge invariance and is deduced directly from
canonical quantization\cite{soldati}. The presence of $\bar{n}_{\mu}$ destroys
the $Sim(2)$ symmetry, though. In order to restore the $Sim(2)$ symmetry we devised a systematic way to take $\bar{n}_{\mu} \rightarrow
0$. Then only one null vector survives. We have tested the new infrared
regularization computing the one loop renormalization of VSR Quantum
electrodynamics with a non-zero photon mass $m_{\gamma}$. A remarkable
contribution to the anomalous magnetic moment of the electron appears,
depending on the electron neutrino and photon mass. It fits very well between
the bounds of the most recent measurements.

Using this infrared regularization of $d$ dimensional integrals, we showed how to recover the expected Yukawa potential in VSR QED \cite{jaren}.

The existence of unphysical potentials in the NRL of VSR theories with gauge invariant mass for the gauge field appears to be generic. It is also present in the gravitational potential in Very Special linear gravity(VSLG)\cite{ASan,ASanS}.The procedure presented in \cite{jaren}  remove the unphysical components of the gravitational potential.

In this paper we consider light by light scattering in VSR Quantum
electrodynamics with a non-zero photon mass. Again, the new regularization
produces amplitudes that preserve gauge invariance and have manifest unitarity. We
evaluate the total cross section $\sigma$. The corrections due to the photon
mass are tiny. Then we give an analytic expression of $\sigma$ including up to
the leading $\frac{m_{\gamma}^2}{\omega^2} \log \left(
\frac{m_{\gamma}^2}{\omega^2} \right)$ term , $m_{\gamma} < \omega < M_e$,
where $\omega$ is the photon energy and $M_e$ is the electron mass. The
leading correction is anisotropic, revealing the loss of rotational symmetry
of VSR. The anisotropy could be observable at cosmological scales. In fact since 20 years ago, various anomalies in the Cosmic Microwave Background radiation (CMB) has been detected \cite{Planck} and strong evidence of cosmological anisotropies is accumulating\cite{Kester}.

Furthermore, we compute the one loop amplitude for $2N$ external photon legs without external electron legs.
They turns out to be equal to the QED's ones with the electron mass $M_e=\sqrt{M^2+m^2}$, as expected from unitarity.
It follows that the Euler-Heisenberg Lagrangian coincides with the one in QED.

The structure of the paper is as follows: Chapter 2 reviews the new infrared regularization defined in \cite{jaren}.
Chapter 3 introduces VSR QED with a gauge invariant photon mass. In chapter 4 we consider light by light scattering in VSR QED with a massive photon.
Since VSR QED reduces to QED for large photon momenta, we prefer to study the scattering process for low photon momenta $k_i$, $k_i^2<<M_e^2$. This permits
to use the Euler-Heisenberg Lagrangian. Chapter 5 contains the average over photon polarizations, to get the unpolarized total cross section. However, we get a much simpler analytic result for the total cross section. In chapter 6, we consider  $N$ photons scattering. In chapter 7 we draw the conclusions.

The Feynman rules where derived in {\cite{plb}}. We list them in Appendix A.In Appendix B we write  the unpolarized probability for photon-photon scattering, keeping up to order $m_\gamma^2$ for simplicity , although we computed it exactly. In Appendix C we generalized an identity considered in \cite{IT}, to include massive photons.

\section{$Sim(2)$ invariant regularization}
Here we review the regularization procedure introduced in \cite{jaren}.

Consider an arbitrary function $g$ and compute
\[ \int \frac{d^d q}{(2 \pi)^d} g (q^2, q.x) \frac{1}{(n.q)^a}\]
The M-L prescription using the method of $\cite{AML}$ implies
\[ \int \frac{d^d q}{(2 \pi)^d} g (q^2, q.x) \frac{1}{(n.q)^a} = (\bar{n}
.x)^a f (x.x, n.x \bar{n} .x) \]
for a unique $f (x.x, n.x \bar{n} .x)$, under the conditions:
\begin{enumerate}
	\item $n.n = 0 = \bar{n} . \bar{n}$, $n. \bar{n} = 1$
	
	\item Scale invariance under $n_{\mu} \rightarrow \lambda n_{\mu},
	\bar{n}_{\mu} \rightarrow \lambda^{- 1} \bar{n}_{\mu}$.
	
	\item $f (x.x, n.x \bar{n}.x)$ must be regular at $n.x \bar{n}.x = 0$.
\end{enumerate}
$x_{\mu}$ is an arbitrary vector.

To take the limit $\bar{n}_{\mu} \rightarrow 0$, write $\bar{n}_{\mu} = \rho
\bar{n}_{\mu}^{(0)}, n_{\mu} = \rho^{- 1} n_{\mu}^{(0)}$, with

$\bar{n}_{\mu}^{(0)}, n_{\mu}^{(0)}$ satisfying condition 1.Then condition 1.
is satisfied for all $\rho$.

We define $\bar{n}_{\mu} \rightarrow 0$ by the limit $\rho \rightarrow 0$.

We get: $\lim_{\rho \rightarrow 0} \rho^a (\bar{n}^{(0)} .x)^a f (x.x, n^{(0)}
.x \bar{n}^{(0)} .x) = 0$.

Thus:
\[ \int \frac{d^d q}{(2 \pi)^d} g (q^2, q.x) \frac{1}{(n.q)^a} = 0, a > 0 \]
It is clear that this result applies to loop integrals of the sort
{\cite{AML}}:
\begin{eqnarray}
	\int dp \frac{1}{[p^2 + 2 p.q - m^2 + i \varepsilon]^a}  \frac{1}{(n \cdot
		p)^b} = (- 1)^{a + b} i (\pi)^{\omega}  (- 2)^b \frac{\Gamma (a + b -
		\omega)}{\Gamma (a) \Gamma (b)}  (\bar{n} \cdot q)^b &  &  \nonumber\\
	\int_0^1 dtt^{b - 1}  \frac{1}{(m^2 + q^2 - 2 n \cdot q \bar{n} \cdot qt - i
		\varepsilon)^{a + b - \omega}}, & \omega = d / 2 &  \label{basic}
\end{eqnarray}
Therefore, the $Sim(2)$ limit is:
\[ \int dp \frac{1}{[p^2 + 2 p.q - m^2 + i \varepsilon]^a}  \frac{1}{(n \cdot
	p)^b} = 0, b > 0, q_{\mu} arbitrary \]

Taking derivatives in $q_{\mu}$, we get:
\[ \int dp \frac{1}{[p^2 + 2 p.q - m^2 + i \varepsilon]^a}  \frac{p_{\alpha_1}
	\ldots .p_{\alpha_n}}{(n \cdot p)^b} = 0, b > 0, q_{\mu} arbitrary
\]
Resuming, the $Sim(2)$ invariant regularization of any integral over
$p_{\mu}$, containing $\frac{1}{n.p}$ to any positive power must be evaluated
to zero.

It is clear that this procedure respects gauge invariance and $Sim(2)$ invariance.	

But how to proceed if $\gamma$ matrices are involved?

Let us consider an example:
\begin{eqnarray*}
	\int \frac{dp}{p^2 - m^2} \frac{\not{n}  \not{p}  \not{n}}{n.p} =
	\int \frac{dp}{p^2 - m^2} \frac{\left( 2 n.p - \not{p \not{n}}
		\right)  \not{n}}{n.p} = 2 \int \frac{dp}{p^2 - m^2} &  & 
\end{eqnarray*}
I can compute $p$ integral first, using ML:
\begin{eqnarray*}
	\int \frac{dp}{p^2 - m^2} \frac{\not{n}  \not{p}  \not{n}}{n.p} =
	\not{n} \gamma_{\mu}  \not{n} \int \frac{dp}{p^2 - m^2}
	\frac{p_{\mu}}{n.p} = \not{n} \gamma_{\mu}  \not{n} \int
	\frac{dp}{p^2 - m^2} \bar{n}_{\mu} &  & 
\end{eqnarray*}
the naive limit will give zero, but if we move $\not{n}$ to the right(or left) $\not{n}  \not{\bar{n}}  \not{n} = 2$ and
we get the same answer as before.

But, suppose that $\not{n}$ was already to the right. Consider:
\begin{eqnarray*}
	\int \frac{dp}{p^2 - m^2} \frac{ \not{p}  \not{n}}{n.p} = \int
	\frac{dp}{p^2 - m^2}  \not{\bar{n}}  \not{n} \rightarrow ? &  & 
\end{eqnarray*}
\emph{\bf Prescription:}
We move all $\not{n}$ to the right, pick up all $n.(p+Q)$ produced by this motion and use them to cancel as many $n.(p+Q)$ in the denominator as possible. Finally all remaining  $(n.(p+Q))^{-a},a>0$ are replaced by zero. Here $Q_\mu$ represents any vector different from $p_\mu$ (the integration variable) including the zero vector. Notice that $\frac{n.p}{n.(p+q)}=1$ because $\frac{n.p}{n.(p+q)}=1-\frac{n.q}{n.(p+q)}$ and the second term vanishes in the last step of the procedure.

According to this:
\begin{eqnarray*}
	\int \frac{dp}{p^2 - m^2} \frac{ \not{p}  \not{n}}{n.p} =0 
\end{eqnarray*}

and
\begin{eqnarray*}
	\int \frac{dp}{p^2 - m^2} \frac{\not{n}  \not{p}  \not{n}}{n.p} =
	2 \int \frac{dp}{p^2 - m^2} &  & 
\end{eqnarray*}

The rationale for this prescription is the following. We are interested in putting $\bar{n}_\mu=0$ to recover $Sim(2)$ invariance. But we cannot afford to loose gauge invariance. Gauge invariance appears in the form of Ward identities that the Feynman graphs must satisfy. If we write all graphs in a "canonical form" such as all $\not{n}$ to the right in all monomials(only one $\not{n}$ remains because $\not{n}.\not{n}=0$ ),the Ward identities that generally involves products with external momenta, will be satisfied for arbitrary values of $n_\mu$ and $\bar{n}_\mu$(To prove the Ward identity we do not need $n.n=\bar{n}.\bar{n}=0$,$n.\bar{n}=1$  anymore when all $\not{n}$ are to the right of all $\not{\bar{n}}$ ). Then after evaluating $\bar{n}_\mu=0$, the Ward identity still will be satisfied in the surviving set of integrals defining the  graphs. This surviving set  defines the $Sim(2)$ invariant gauge theory.

The prescription has a degree of arbitrariness. We could equally well use the convention of moving all $\not{n}$ to the left.

In the application to VSR QED we have checked whether this arbitrariness in the prescription produces ambiguities. We did not find any.

\section{The model}

The leptonic sector of VSRSM consists of three $SU (2)$ doublets $L_{a} =
\left( \begin{array}{c}
	\nu^{0}_{aL}\\
	e^{0}_{aL}
\end{array} \right)$, where $\nu^{0}_{aL} = \frac{1}{2}  (1- \gamma_{5} )
\nu^{0}_{a}$ and $e^{0}_{aL} = \frac{1}{2}  (1- \gamma_{5} ) e^{0}_{a}$, and
three $SU (2)$ singlet $R_{a} =e^{0}_{aR} = \frac{1}{2}  (1+ \gamma_{5} )
e^{0}_{n}$. We assume that there is no right-handed neutrino. The index $a$
represent the different families and the index $0$ says that the fermionic
fields are the physical fields before breaking the symmetry of the vacuum.

In this work we restrict ourselves to the electron family.

$m$ is the VSR mass of both electron and neutrino. 

After spontaneous symmetry breaking(SSB), the electron acquires a mass term
$M= \frac{G_{e} v}{\sqrt{2}}$, where $G_{e}$ is the electron Yukawa coupling
and $v$ is the VEV of the Higgs. Please see equation (52) of {\cite{ja1}}. The electron mass is
$M_e = \sqrt{M^2 + m^2}$.The neutrino mass is not affected by SSB:$M_{\nu_e}= m$.

Restricting the VSRSM after SSB to photon and electron alone, we get the VSR
QED action with a massive photon.$\psi$:electron field. $A_{\mu}$:photon
field. $m_{\gamma}$:photon mass. We use Feynman gauge.
\begin{eqnarray}
	\mathcal{L} = \bar{\psi}  \left( i \left( \not{D} + \frac{1}{2} \not{n} m^2
	(n \cdot D)^{- 1} \right) - M \right) \psi - \frac{1}{4} F_{\mu \nu} F^{\mu
		\nu} &  & \nonumber\\
	\hspace{-1cm}- \frac{1}{2} m_{\gamma}^2  (n^{\alpha} F_{\mu \alpha}) \frac{1}{(n \cdot
		\partial)^2}  (n_{\beta} F^{\mu \beta}) - \frac{(\partial_{\mu}
		A_{\mu})^2}{4} 
\end{eqnarray}
where $D_{\mu} = \partial_{\mu} - ieA_{\mu}, F_{\mu \nu} = \partial_{\mu} A_{\nu} -
	\partial_{\nu} A_{\mu}$ and $n.n = 0$.

This Lagrangian (without the gauge fixing term) is gauge invariant under the
usual gauge transformations:$\delta A_{\mu} (x) = \partial_{\mu} \Lambda (x)$.
This is a very important property of a massive photon in VSR. It preserves
gauge invariance. Instead a Lorentz invariant mass for the photon breaks gauge
invariance.

\section{Photon-photon scattering in VSR QED}
The Feynman rules are given in Appendix A. The graphs that contribute to photon-photon scattering in VSR QED are contained in Figure 1.To draw the graphs we used \cite{ellis}.
\begin{figure}[h]
	\includegraphics[scale=0.4]{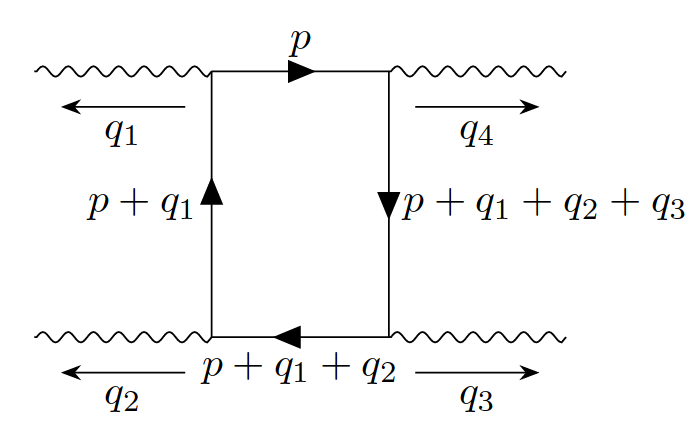}
	\includegraphics[scale=0.4]{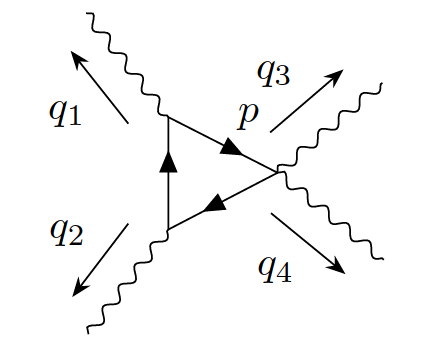}
	\includegraphics[scale=0.4]{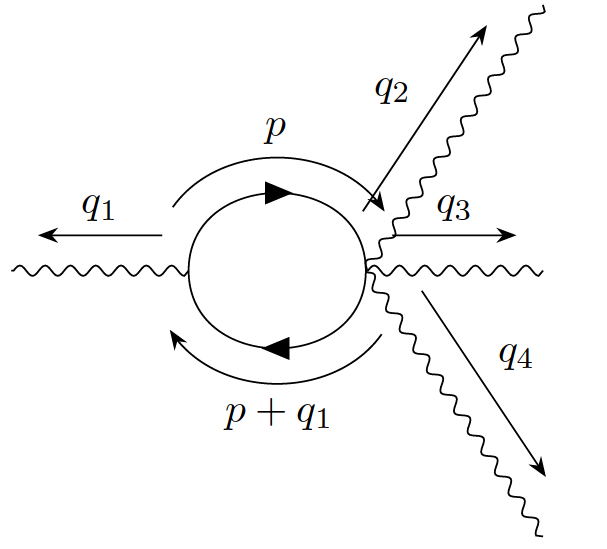}
	\caption{One loop graphs for light by light scattering in VSR QED.$\sum_i q_i = 0$}
	\label{Fig: photon-photon scattering}
\end{figure}

It is easy to check that the infrared regularization implies that graphs (2,3)
vanish, because the negative powers of $n.p$ are too large to be compensated by positive powers generated by the trace.

Define
$V_{\mu} (p', p) = \left( \gamma_{\mu} + \frac{1}{2} m^2 \frac{\not{n}}{n.p
	n.p'} n_{\mu} \right)$, $S_F (p) = \frac{\not{p} + M - \frac{m^2}{2}
	\frac{\not{n}}{n.p}}{p^2 - M_e^2 + i \varepsilon}$

We illustrate this  using graph (2). 
\begin{eqnarray*}
	C_{\mu_1 \mu_2 \mu_3 \mu_4} = &  & \\
	- (- i e^2)^3 \frac{1}{2} n_{\mu_3} n_{\mu_4} m^2 \int d p Tr \left\{
	S_F (p) V_{\mu_1} (p + q_1, p) S_F (p + q_1) V_{\mu_2} (p + q_1 + q_2, p +
	q_1) S_F (p + q_1 + q_2) \not{n} \right\}\\ \frac{1}{n. (p + q_1 + q_2)}
	\frac{1}{n.p} \left( \frac{1}{n. (q_3 + p)} + \frac{1}{n. (q_4 + p)} \right)
	&  & 
\end{eqnarray*}
The $Sim(2)$ symmetry implies the same number of $n_{\mu}$ in the
numerator as in the denominator. But due to the new VSR vertex, two
$n_{\mu}$'s are outside the integral. In the denominator we have 3 $n.p$ type
factors, so the trace will produce at most one $n.p$ factor. The integral vanishes. 

The same reasoning shows that the graph (3) vanishes.

Moreover graph (1) reduces to the standard QED result with electron
mass given by $M_e = \sqrt{M^2 + m^2}$. In this way unitarity is recovered. That is:
\begin{eqnarray*}
	- (- i e)^4 \int \frac{d^n p}{(2 \pi)^n} Tr \{ V_{\mu_1} S_F (p +
	q_1) V_{\mu_2} S_F (p + q_1 + q_2) V_{\mu_3} S_F (p + q_1 + q_2 + q_3)
	V_{\mu_4} S_F (p) \} = &  & \\
	- (- i e)^4 \int \frac{d^n p}{(2 \pi)^n} Tr\{ \gamma_{\mu_1} S'_F (p
	+ q_1) \gamma_{\mu_2} S'_F (p + q_1 + q_2) \gamma_{\mu_3} S'_F (p + q_1 +
	q_2 + q_3) \gamma_{\mu_4} S'_F (p) \} &  & 
\end{eqnarray*}
with $S'_F (p) = \frac{i \left( \not{p} + M_e \right)}{p^2 - M^2_e + i\varepsilon}$

In any step of the regularization procedure, we are using dimensional regularization. Therefore the amplitude for photon-photon scattering is gauge invariant \cite{Khare}

We want to compute the scattering cross section for this process.

\section{Photon-Photon scattering for $q_i^2 \ll M_e^2$}

The exact result for the amplitude  is complicated {\cite{Landau}}.

But since the mass of the photon is very small, we will not have any
difference from QED, except in the $k_i \rightarrow 0$ limit.$k_i^2 \ll
M_e^2$. But in this region the amplitude reduces to the Euler-Heisenberg
Lagrangian, which is much simpler to write.

Keeping up to $o (F_{\mu \nu}^4)$ which describes photon-photon scattering we
get{\cite{IT}}:
\begin{eqnarray}
	\mathcal{L}= -\mathcal{F}+ \frac{\alpha^2}{360}  \frac{1}{M_e^4}  (4 (F_{\mu
		\nu} F^{\mu \nu})^2 + 7 (F_{\mu \nu}  \tilde{F}^{\mu \nu})^2) &  & 
	\label{EH}
\end{eqnarray}
where: $\mathcal{F}= \frac{1}{4} F_{\mu \nu} F^{\mu \nu}$, $\tilde{F}_{\mu
	\nu} = \frac{1}{2} \varepsilon_{\mu \nu \alpha \beta} F^{\alpha \beta}$;
$\alpha$ is the fine structure constant.

Using equation(\ref{EH}) is very simple to derive the photon-photon scattering
amplitude for $q_i^2 \ll M_e^2$ {\cite{IT}}. It is:
\begin{eqnarray}
	M = - \frac{i \alpha^2}{45 M_e^2}\frac{1}{12}\nonumber\\   (5 (trf_1 f_2 trf_3 f_4 +
	trf_1 f_3 trf_2 f_4 + trf_1 f_4 trf_2 f_3)  \nonumber\\
	- 7 tr (f_1 f_2 f_3 f_4 + f_2 f_1 f_3 f_4 + f_3 f_1 f_2 f_4 +\nonumber\\ f_2 f_3 f_1
	f_4 + f_3 f_2 f_1 f_4 + f_3 f_2 f_1 f_4 + f_1 f_3 f_2 f_4))
\end{eqnarray}
with
$f_{i \rho \sigma} = i (q_{i \rho} \epsilon_{i \sigma} - q_{i \sigma}
\epsilon_{i \rho})$

\section{VSR Cross section with unpolarized photons}

Notice that the amplitude is formally equal to the QED result. The VSR
contribution is hidden in the vectors $\epsilon_{i\sigma}$. Also we have $q_{i\mu}q_i^\mu=m_\gamma^2$. The $Sim(2)$ symmetry allows to write $n_{\mu} = (1, \hat{n})$, $\hat{n} . \hat{n} =
1$. For light by light scattering we take $q_1=k_1,q_2=k_2,q_3=-k_3,q_4=-k_4$.

The sum over  polarizations in VSR is given by\cite{AS1}:
\begin{eqnarray*}
	\sum_{\lambda} \epsilon_{\mu} \epsilon^{\ast}_{\nu} = - g_{\mu \nu} -
	\frac{m_{\gamma}^2}{(n \cdot k)} n_{\mu} n_{\nu} &  & 
\end{eqnarray*}
The unpolarized differential cross section is given by\cite{IT}:
\begin{eqnarray}
	\frac{d \sigma}{d \Omega} = \frac{1}{(2 \pi)^2} \frac{1}{2 \omega^2}
	\frac{\alpha^4}{(90)^2 M_e^8} P
\end{eqnarray}
$P$ is written in Appendix B.

The total cross section is:
\begin{eqnarray*}
	\sigma = \frac{1}{(2 \pi)^2} \frac{1}{2 \omega^2} \frac{\alpha^4}{(90)^2
		M_e^8} \int d \Omega P &  & 
\end{eqnarray*}
To compute the total cross section $\sigma$ we work in the CM system. Most of the calculations have used FORM \cite{form}. Then:
\begin{eqnarray*}
	k_1 = (\omega, \vec{k}), &  & k_2 = (\omega, - \vec{k}),\\
	k_3 = (\omega, \vec{k}'), &  & k_2 = (\omega, - \vec{k}')
\end{eqnarray*}
To integrate over the solid angle, we choose polar coordinates for $\vec{k}'$
with the $z$-axis in the direction of $\hat{n}$. Keeping up to $o \left(
\frac{m_{\gamma}^2}{\omega^2} \log \left( \frac{m_{\gamma}^2}{\omega^2}
\right) \right)$ and including the factor $\frac{1}{2}$ due to Bose
statistics, we get:
\begin{equation}
\sigma = \frac{973}{10125 \pi} \frac{\alpha^4}{M_e^8} \omega^6 \left( 1 -
\frac{5}{56}  \frac{m_{\gamma}^2}{\omega^2} \hspace{0.17em} \log \left(
\frac{4 \omega^2}{m_{\gamma}^2} \right) (3 + \cos^2 \alpha)^2 \right)	
\end{equation}
where
$\vec{k} . \hat{n} = k \cos \alpha$. The result holds for $m_{\gamma} < \omega < M $

The leading correction is anisotropic, revealing the loss of rotational
symmetry of VSR.

From Particle Data Group{\cite{pdg}}:
\begin{equation}
	m_{\gamma} < 3 \times 10^{- 27} eV / c^2
\end{equation}

For extremely low frequency radio waves(ELF), $\omega\sim 10^{-14}$ eV,corresponding to a wavelength $\lambda\sim 10^8$m.  so the anisotropic term is very small and no conflict with present experimental data appears.

To detect the anisotropy we have to look at cosmological scales.
In fact we expect that our result will produce tiny but measurable anisotropies in the Cosmic Microwave Background Radiation (CMB).

CMB anomalies have been pointed out since 20 years ago \cite{Planck} and strong evidence of a dipole cosmic anisotropy has been accumulating\cite{Kester}.

In \cite{ja1} we already speculated about the cosmic origin of 
the privileged direction in VSR, given by $\vec{n}$.

It is enticing to think that the physical cause of the small neutrinos masses and tiny photon and graviton masses is due to a primordial dipole anisotropy of the Universe.

\section{$2N$ legs and Euler-Heisenberg Lagrangian}

The same reasoning from section 4 applies to  6,8... $2N$photon graphs.

We reach the following conclusions:

1)Additional VSR graphs, obtained by the insertion of 2,3,..$2N-1$ vertices's, vanish in the $Sim(2)$ limit. In fact
the $Sim(2)$ symmetry implies the same number of $n_{\mu}$ in the
numerator as in the denominator.  But  insertions of  the new VSR vertices's  imply that
$n_{\mu}$'s are outside the integral. Then the number of $n.p$ type
factors that is integrated over is negative. Therefore the integral vanishes in the $\bar{n}_\mu=0$ limit.

2)The remaining graphs corresponding to standard QED graphs reduce to the
standard QED result with $M_e^2 = M^2 + m^2$ as required by unitarity.

Therefore  the $\bar{n}_{\mu} \rightarrow 0$ limit of the Euler-Heisenberg
Lagrangian in VSR coincides with the E-H lagrangian in QED with $M_e^2 = M^2 +m^2$.

To obtain conclusion 2), use the  $\not{n}$ to the right prescription and replace $m^2=M_e^2-M^2$. By explicit computation for $N=1,..,4$, we found that  $M$ appears to the first power alone. The coefficient of M is made of functions of $M_e^2$ multiplied by an odd number of Dirac's matrices. So their trace vanishes. This seems to hold for all $N$.

\section{Conclusions}

We have computed the unpolarized differential cross section for light by light scattering in VSR QED with a gauge invariant photon mass, in terms of Lorentz scalar functions (Appendix B). We restricted our calculation to $\vec{k}_i^2< M_e^2$, since for higher values of the momenta, VSR QED coincides with QED. Using this result we have obtained the leading contribution in the limit $m_\gamma\rightarrow 0$ to the total cross section $\sigma$  in the CM system. The leading contribution is anisotropic, exhibiting the loss of rotational invariance of VSR models. But the photon mass term is very small, so no conflict with available experimental data appears.

In fact we expect that our result will produce tiny but measurable anisotropies in the Cosmic Microwave Background Radiation (CMB). The  anisotropy is due to a preferred direction $\hat{n}$.

We apply the infrared regulator to one loop amplitudes with an arbitrary number of photon legs, reaching the ensuing conclusions:

1)Additional VSR graphs, obtained by the insertion of 2,3,..$2N-1$ vertices's, vanish in the $Sim(2)$ limit.

2)The remaining graphs corresponding to standard QED graphs reduce to the
standard QED result with $M_e^2 = M^2 + m^2$ as required by unitarity.

Therefore  the $\bar{n}_{\mu} \rightarrow 0$ limit of the Euler-Heisenberg
Lagrangian in VSR coincides with the E-H lagrangian in QED with $M_e^2 = M^2 +
m^2$ as required by unitarity.

It is remarkable how the new infrared regulator simplifies the computation of all VSR graphs, reducing them to standard dimensionally regularized integrals,
all along preserving gauge invariance and $Sim(2)$ symmetry. The way is open to explore the contribution of massive neutrinos and massive photon in more complicated but interesting settings.

\begin{section}{Acknowledgments}
	
	J.A.acknowledges the partial support of the Institute of
	Physics PUC.
\end{section}

\section{Appendix A:Feynman rules}
To draw the Feynman graphs we used \cite{ellis}
\begin{figure}[h]
	\includegraphics[scale=0.4]{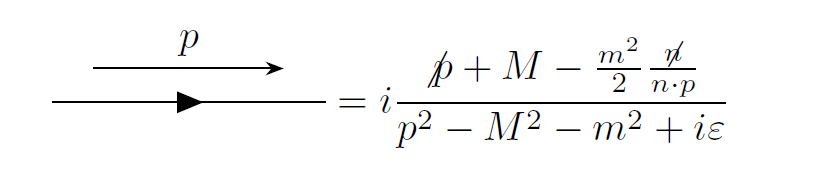}
	\includegraphics[scale=0.4]{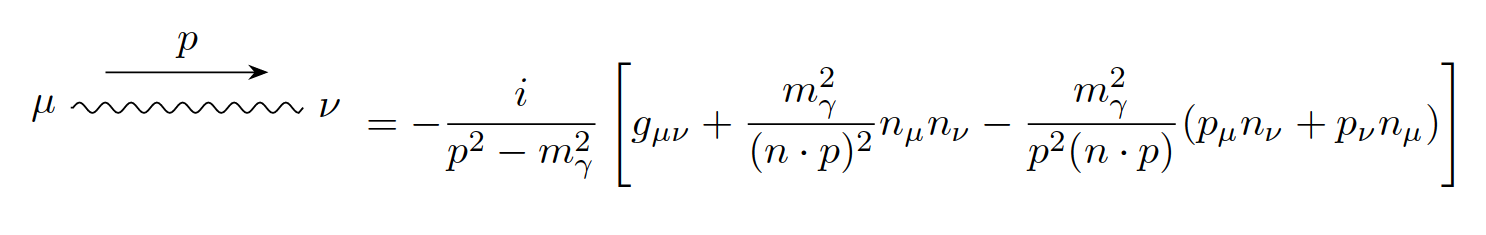}
	\includegraphics[scale=0.4]{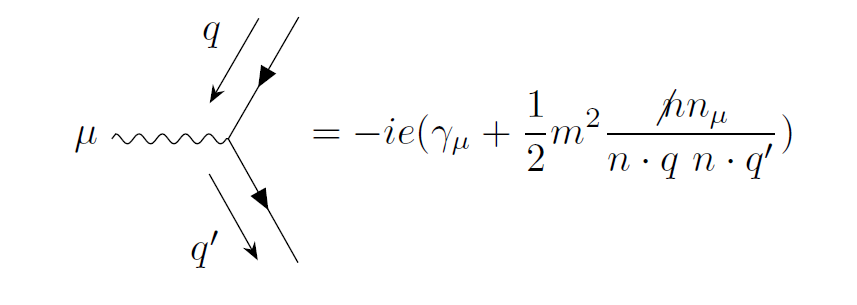}
	\includegraphics[scale=0.4]{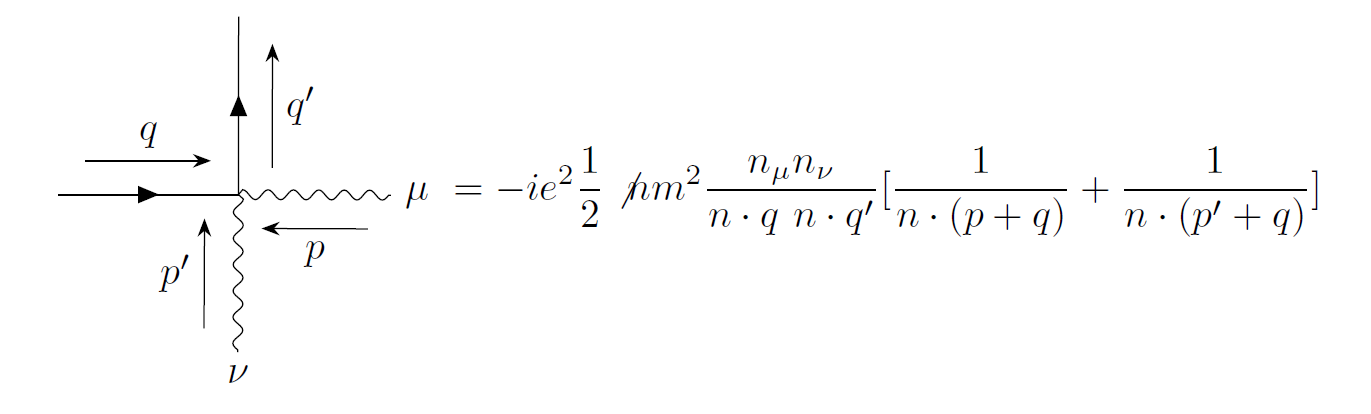}
	\caption{Feynman rules for one loop computations:electron propagator,photon propagator,$A_\mu ee$ and $A_\mu A_\nu ee$ vertex.}
	\label{Fig: Fey rules}
\end{figure}	
\section{Appendix B:Unpolarized probability in Form notation up to $o(m_\gamma^2)$}

\begin{verbatim}
P=
 mg^2 * (  - 321*n.k1^-2*n.k2*n.k3*k1.k2*k1.k4*k3.k4 - 321*n.k1^-2*
n.k2*n.k3*k1.k3*k1.k4*k2.k4 + 321*n.k1^-2*n.k2*n.k3*k1.k4^2*k2.k3 - 
321*n.k1^-2*n.k2*n.k4*k1.k2*k1.k3*k3.k4 - 321*n.k1^-2*n.k2*n.k4*k1.k3
*k1.k4*k2.k3 + 321*n.k1^-2*n.k2*n.k4*k1.k3^2*k2.k4 + 321*n.k1^-2*
n.k2^2*k1.k3*k1.k4*k3.k4 - 321*n.k1^-2*n.k3*n.k4*k1.k2*k1.k3*k2.k4 - 
321*n.k1^-2*n.k3*n.k4*k1.k2*k1.k4*k2.k3 + 321*n.k1^-2*n.k3*n.k4*
k1.k2^2*k3.k4 + 321*n.k1^-2*n.k3^2*k1.k2*k1.k4*k2.k4 + 321*n.k1^-2*
n.k4^2*k1.k2*k1.k3*k2.k3 + 164*n.k1^-1*n.k2*k1.k2*k3.k4^2 - 303*
n.k1^-1*n.k2*k1.k3*k2.k4*k3.k4 - 303*n.k1^-1*n.k2*k1.k4*k2.k3*k3.k4
- 303*n.k1^-1*n.k3*k1.k2*k2.k4*k3.k4 + 164*n.k1^-1*n.k3*k1.k3*
k2.k4^2 - 303*n.k1^-1*n.k3*k1.k4*k2.k3*k2.k4 - 303*n.k1^-1*n.k4*k1.k2
*k2.k3*k3.k4 - 303*n.k1^-1*n.k4*k1.k3*k2.k3*k2.k4 + 164*n.k1^-1*n.k4*
k1.k4*k2.k3^2 - 321*n.k1*n.k2^-2*n.k3*k1.k2*k2.k4*k3.k4 + 321*n.k1*
n.k2^-2*n.k3*k1.k3*k2.k4^2 - 321*n.k1*n.k2^-2*n.k3*k1.k4*k2.k3*k2.k4
- 321*n.k1*n.k2^-2*n.k4*k1.k2*k2.k3*k3.k4 - 321*n.k1*n.k2^-2*n.k4*
k1.k3*k2.k3*k2.k4 + 321*n.k1*n.k2^-2*n.k4*k1.k4*k2.k3^2 + 164*n.k1*
n.k2^-1*k1.k2*k3.k4^2 - 303*n.k1*n.k2^-1*k1.k3*k2.k4*k3.k4 - 303*n.k1
*n.k2^-1*k1.k4*k2.k3*k3.k4 + 321*n.k1*n.k2*n.k3^-2*k1.k2*k3.k4^2 - 
321*n.k1*n.k2*n.k3^-2*k1.k3*k2.k4*k3.k4 - 321*n.k1*n.k2*n.k3^-2*k1.k4
*k2.k3*k3.k4 + 321*n.k1*n.k2*n.k4^-2*k1.k2*k3.k4^2 - 321*n.k1*n.k2*
n.k4^-2*k1.k3*k2.k4*k3.k4 - 321*n.k1*n.k2*n.k4^-2*k1.k4*k2.k3*k3.k4
- 321*n.k1*n.k3^-2*n.k4*k1.k2*k2.k3*k3.k4 - 321*n.k1*n.k3^-2*n.k4*
k1.k3*k2.k3*k2.k4 + 321*n.k1*n.k3^-2*n.k4*k1.k4*k2.k3^2 - 303*n.k1*
n.k3^-1*k1.k2*k2.k4*k3.k4 + 164*n.k1*n.k3^-1*k1.k3*k2.k4^2 - 303*n.k1
*n.k3^-1*k1.k4*k2.k3*k2.k4 - 321*n.k1*n.k3*n.k4^-2*k1.k2*k2.k4*k3.k4
+ 321*n.k1*n.k3*n.k4^-2*k1.k3*k2.k4^2 - 321*n.k1*n.k3*n.k4^-2*k1.k4*
k2.k3*k2.k4 - 303*n.k1*n.k4^-1*k1.k2*k2.k3*k3.k4 - 303*n.k1*n.k4^-1*
k1.k3*k2.k3*k2.k4 + 164*n.k1*n.k4^-1*k1.k4*k2.k3^2 + 321*n.k1^2*
n.k2^-2*k2.k3*k2.k4*k3.k4 + 321*n.k1^2*n.k3^-2*k2.k3*k2.k4*k3.k4 + 
321*n.k1^2*n.k4^-2*k2.k3*k2.k4*k3.k4 - 321*n.k2^-2*n.k3*n.k4*k1.k2*
k1.k3*k2.k4 - 321*n.k2^-2*n.k3*n.k4*k1.k2*k1.k4*k2.k3 + 321*n.k2^-2*
n.k3*n.k4*k1.k2^2*k3.k4 + 321*n.k2^-2*n.k3^2*k1.k2*k1.k4*k2.k4 + 321*
n.k2^-2*n.k4^2*k1.k2*k1.k3*k2.k3 - 303*n.k2^-1*n.k3*k1.k2*k1.k4*k3.k4
- 303*n.k2^-1*n.k3*k1.k3*k1.k4*k2.k4 + 164*n.k2^-1*n.k3*k1.k4^2*
k2.k3 - 303*n.k2^-1*n.k4*k1.k2*k1.k3*k3.k4 - 303*n.k2^-1*n.k4*k1.k3*
k1.k4*k2.k3 + 164*n.k2^-1*n.k4*k1.k3^2*k2.k4 - 321*n.k2*n.k3^-2*n.k4*
k1.k2*k1.k3*k3.k4 - 321*n.k2*n.k3^-2*n.k4*k1.k3*k1.k4*k2.k3 + 321*
n.k2*n.k3^-2*n.k4*k1.k3^2*k2.k4 - 303*n.k2*n.k3^-1*k1.k2*k1.k4*k3.k4
- 303*n.k2*n.k3^-1*k1.k3*k1.k4*k2.k4 + 164*n.k2*n.k3^-1*k1.k4^2*
k2.k3 - 321*n.k2*n.k3*n.k4^-2*k1.k2*k1.k4*k3.k4 - 321*n.k2*n.k3*
n.k4^-2*k1.k3*k1.k4*k2.k4 + 321*n.k2*n.k3*n.k4^-2*k1.k4^2*k2.k3 - 303
*n.k2*n.k4^-1*k1.k2*k1.k3*k3.k4 - 303*n.k2*n.k4^-1*k1.k3*k1.k4*k2.k3
+ 164*n.k2*n.k4^-1*k1.k3^2*k2.k4 + 321*n.k2^2*n.k3^-2*k1.k3*k1.k4*
k3.k4 + 321*n.k2^2*n.k4^-2*k1.k3*k1.k4*k3.k4 + 321*n.k3^-2*n.k4^2*
k1.k2*k1.k3*k2.k3 - 303*n.k3^-1*n.k4*k1.k2*k1.k3*k2.k4 - 303*n.k3^-1*
n.k4*k1.k2*k1.k4*k2.k3 + 164*n.k3^-1*n.k4*k1.k2^2*k3.k4 - 303*n.k3*
n.k4^-1*k1.k2*k1.k3*k2.k4 - 303*n.k3*n.k4^-1*k1.k2*k1.k4*k2.k3 + 164*
n.k3*n.k4^-1*k1.k2^2*k3.k4 + 321*n.k3^2*n.k4^-2*k1.k2*k1.k4*k2.k4 + 
285*k1.k2*k1.k3*k2.k3 + 285*k1.k2*k1.k4*k2.k4 + 285*k1.k3*k1.k4*k3.k4
+ 285*k2.k3*k2.k4*k3.k4 )

+ 139*k1.k2^2*k3.k4^2 + 139*k1.k3^2*k2.k4^2 + 139*k1.k4^2*k2.k3^2
\end{verbatim}
\section{Appendix C:Scalar identity for massive photon}
We write below the generalization of the identity used in \cite{IT} to simplify (7-99). Notice that the matrix $k_i.k_j$ has a eigenvector with eigennvalue zero, $(1,1,-1,-1)$, corresponding to 4-momentum conservation.
\begin{verbatim}
       Det(ki.kj)=0= mg^2 * ( k1.k2*k1.k3*k2.k3 + k1.k2*k1.k4*k2.k4 + k1.k3*k1.k4*k3.k4 + 
k2.k3*k2.k4*k3.k4 )

+ mg^4 * (  - 1/2*k1.k2^2 - 1/2*k1.k3^2 - 1/2*k1.k4^2 - 1/2*k2.k3^2 - 1/
2*k2.k4^2 - 1/2*k3.k4^2 )

+ mg^8 * ( 1/2 )

- k1.k2*k1.k3*k2.k4*k3.k4 - k1.k2*k1.k4*k2.k3*k3.k4 + 1/2*k1.k2^2*
k3.k4^2 - k1.k3*k1.k4*k2.k3*k2.k4 + 1/2*k1.k3^2*k2.k4^2 + 1/2*k1.k4^2
*k2.k3^2;
\end{verbatim}

\end{document}